\documentclass[twocolumn, tighten, times]{aastex631}
\usepackage{bm}
\usepackage{amsmath}

\shorttitle{Source identification of KM3-230213A}
\shortauthors{Das et al.}
\graphicspath{{./}{figures/}}

\begin{document}


\title{Cosmic-Ray Constraints on the Flux of Ultra-High-Energy Neutrino Event KM3-230213A

}

\author[0000-0001-5796-225X]{Saikat Das}
\affiliation{Department of Physics, University of Florida, Gainesville, FL 32611, USA}
\email{saikatdas@ufl.edu}

\author[0000-0002-9725-2524]{Bing Zhang}
\affiliation{Nevada Center for Astrophysics, University of Nevada, Las Vegas, Las Vegas, NV 89154, USA}
\affiliation{Department of Physics and Astronomy, University of Nevada, Las Vegas, Las Vegas, NV 89154, USA}

\author[0000-0002-0130-2460]{Soebur Razzaque}
\affiliation{Centre for Astro-Particle Physics (CAPP) and Department of Physics, University of Johannesburg, PO Box 524, Auckland Park 2006, South Africa}
\affiliation{Department of Physics, The George Washington University, Washington, DC 20052, USA}
\affiliation{National Institute for Theoretical and Computational Sciences (NITheCS), Private Bag X1, Matieland, South Africa}

\author[0000-0002-0458-7828]{Siyao Xu}
\affiliation{Department of Physics, University of Florida, Gainesville, FL 32611, USA}




\begin{abstract}

The detection of a $\simeq220$~PeV muon neutrino event by the KM3NeT telescope offers an unprecedented opportunity to probe the Universe at extreme energies. A photopion interaction origin of the neutrino requires a parent cosmic-ray energy of $\gtrsim4$~EeV per nucleon. We analyze the origin of this event under three scenarios, i.e., a transient point source, diffuse astrophysical emission, and a line-of-sight interaction of an ultrahigh-energy cosmic-ray (UHECR; $E\gtrsim 0.1$~EeV).
Our analysis includes the flux from both a KM3NeT-only fit and a joint fit, incorporating data from KM3NeT, IceCube, and the Pierre Auger Observatory.
If the neutrino event originates
from transients, it requires a new population of transients that is energetic, $\gamma$-ray dark, and more abundant
than the known ones.
In the framework of diffuse astrophysical emission, we compare the required local UHECR energy injection rate at $\gtrsim4$ EeV with the rate derived from the flux measurements by Auger, across various source redshift evolution models.
This disfavors the KM3NeT-only fit considering the source evolution up to high values of redshift, while the joint fit remains viable for sources contributing up to a maximum redshift $z_{\rm max} \gtrsim 1$ for the limiting case of photopion interaction efficiency, $f_{p\gamma} = 0.1$. For a cosmogenic origin from point sources, the luminosity obtained at redshifts $z \lesssim 1$ from the joint fit is compatible with the Eddington luminosity of $\sim10^9 M_\odot$ black holes in active galactic nuclei, assuming a proton composition and optimistic values of extragalactic magnetic field strength.
\end{abstract}

\keywords{High energy astrophysics(739) --- Neutrino astronomy(1100) --- Cosmic rays(329) --- Ultra-high-energy cosmic radiation(1733) --- Cosmic ray sources(328)	--- Transient sources(1851)}


\section{Introduction\label{sec:intro}} 

%
The detection of astrophysical neutrinos has opened a unique window to study cosmic accelerators and allows us to probe the universe at distances otherwise inaccessible by $\gamma$ rays and cosmic rays \citep[see, e.g.,][]{Murase:2022feu}. On February 13, 2023, the ARCA detector of KM3NeT, operating for a total time of 287.4 days, recorded an exceptionally 
energetic track-like event, designated KM3-230213A \citep{KM3NeT:2025npi, KM3NeT:2025ccp}. The reconstructed muon energy is estimated to be $120^{+110}_{-60}$~PeV at 68\% C.L., implying a parent neutrino energy of $\sim 220$~PeV, assuming charged-current interaction of a muon neutrino. 

%
The IceCube neutrino observatory has detected an astrophysical neutrino spectrum in the TeV-PeV energy range at $\gtrsim 5\sigma$ significance \citep{IceCube:2013low, IceCube:2014stg}, with 6-13\% of the flux arising in Galactic diffuse emission \citep{IceCube:2023ame}. The detection by KM3NeT significantly extends the observed neutrino energy range beyond the $\sim10$~PeV limit of IceCube's current dataset. The low flux of the Galactic diffuse emission at this energy and zenith angle of $92.1^\circ$ of the KM3-230213A event indicates an extragalactic astrophysical origin \citep{KM3NeT:2025aps}. Cosmological scenarios such as super-heavy dark matter decay, sterile neutrinos, and dark matter scattering have also been explored \citep{Narita:2025udw, Kohri:2025bsn, Choi:2025hqt, Murase:2025uwv, Dev:2025czz}.
If attributed to a diffuse isotropic neutrino flux, the detection is in mild tension with the absence of similar high-energy events in IceCube \citep{KM3NeT:2025ccp,  Li:2025tqf}, based on the KM3NeT-only analysis.

%
A diffuse cosmogenic neutrino flux is expected from the interaction of ultrahigh-energy cosmic rays (UHECRs; $E \gtrsim 10^{17}$ eV) with cosmic background photons \citep[see, e.g.,][]{Kotera_2010}. It has been proposed that a strong redshift evolution of the source population, along with a non-negligible proton fraction in the highest-energy UHECR spectrum, can explain the observed event under the cosmogenic neutrino hypothesis \citep{KM3NeT:2025vut, Zhang:2025abk}. A simultaneous fit to the neutrino data from IceCube and KM3NeT, along with the UHECR spectrum and composition data from the Pierre Auger Observatory \citep{PierreAuger:2023kgv}, suggests a cosmic-ray source population that has not yet been robustly detected by Auger \citep{Muzio:2025gbr}. 

If the current imaging Cherenkov and/or air-shower $\gamma$-ray telescopes detect an associated cosmogenic $\gamma$-ray flux, the extragalactic magnetic field strength can be constrained \citep{Fang:2025nzg}.
While year-long transients \citep{Neronov:2025jfj} and blazars \citep{Dzhatdoev:2025sdi, Yuan:2025zwe} have been proposed as possible sources, no spatial or temporal coincidence of high-energy $\gamma$-ray emission with the neutrino event was observed. Moreover, no statistically significant Fermi-LAT source is identified based on the associated $\gamma$-ray cascade analysis
\citep{Crnogorcevic:2025vou}. 

%
In this article, we consider three plausible origins of the KM3NeT detection. 

(i) First, we estimate the required neutrino energy injection rate at various redshifts for the KM3NeT neutrino flux level. We compare this with that of known transient source populations. 
We analyze different types of $\gamma$-ray burst (GRB) subclasses along with choked jets or failed GRBs, which are also closely related to low-luminosity GRBs. Additionally, tidal disruption events (TDEs), superluminous supernovae (SLSNe), and magnetars are tested as plausible candidates, incorporating their luminosity functions, local event rate densities, as well as their ability to accelerate high-energy cosmic rays to produce the KM3NeT neutrino event. 

(ii) Next, we compute the local UHECR energy injection rate from the observed Auger flux at $\sim4$ EeV, which corresponds to the primary energy per nucleon required for neutrino production via photopion interactions. 
This upper bound constrains the required redshift range of the source distribution, assuming a diffuse astrophysical flux. Redshift evolution models, e.g., star formation rate \citep{Hopkins:2006bw}, luminosity-dependent density evolution (LDDE) of blazars \citep{Ajello_2012, Ajello:2013lka}, and a compact binary merger-rate–based model \citep{Sun:2015bda,Zhu:2020ffa} are tested. 
The $p\gamma$ opacity in these source classes is not well constrained. We present our results for an optimistic interaction efficiency $f_{p\gamma}\sim 0.1$.

(iii) Blazars have been proposed as promising cosmic-ray and neutrino sources. 17 such candidates were found within the 99\% confidence region centered at KM3-230213A \citep{KM3NeT:2025bxl}. In the third scenario, we estimate the kinetic luminosity in jets needed to explain the event via line-of-sight interactions of UHECRs from candidate blazars with known redshifts. Our results constrain the plausible redshift of the neutrino source based on the Eddington luminosity limit, assuming a proton primary and adopting a reasonable extragalactic magnetic field strength as a benchmark case.

We present our results for each of these scenarios in Sections~\ref{sec:transient}, \ref{sec:diffuse}, and \ref{sec:cosmogenic}, respectively. We discuss the implications and conclusions of our results in Sec.~\ref{sec:discussions}.

\section{Transient Neutrino Source\label{sec:transient}}

\begin{figure*}
    \includegraphics[width=0.48\textwidth]{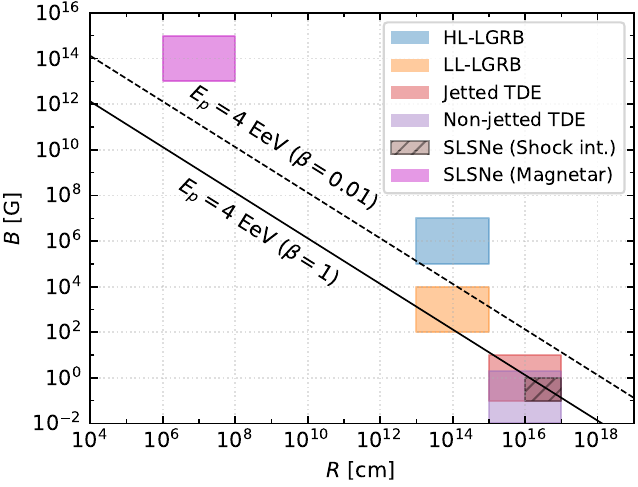}%
    \hfill
    \includegraphics[width=0.48\textwidth]{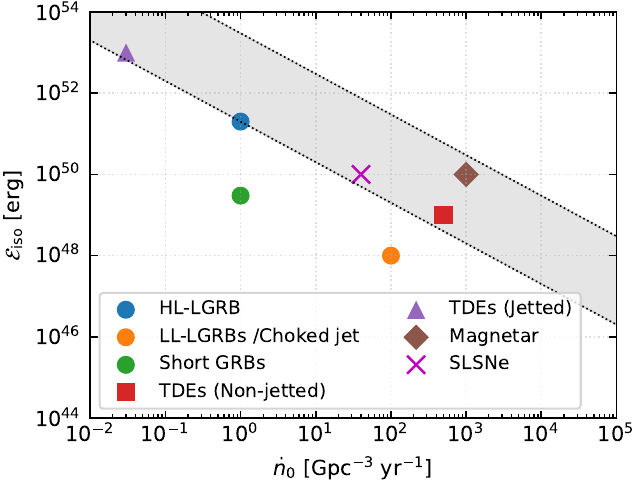}
    \caption{\textit{Left}: Hillas diagram for acceleration of UHE protons to $4$ EeV. The black solid and dashed lines correspond to the velocity of the scattering centers $\beta=v/c=1$ and 0.01, respectively. The shaded boxes show the range of $R$ and $B$ values (listed in Table~\ref{tab:sources}). For relativistic outflows such as GRBs, jetted TDEs, and magnetar-powered SLSNe, the value of Larmor radius $R=\Gamma R^\prime$, where $R^\prime$ is the comoving radius.
    \textit{Right}: Isotropic energy release vs. local event rate density of various candidates. The shaded region represents the estimated electromagnetic energy output associated with the KM3NeT event, assuming it is approximately one order of magnitude lower than the neutrino energy injection rate. The latter is calculated using Eqn.~\ref{eqn:nu_pt} for the joint fit flux level, in the redshift range $z \sim 0.01$--$1$. Representative values are shown for each class to reflect the broader source population, though individual events may vary. The corresponding values are presented in the Table~\ref{tab:sources}.
    }
    \label{fig:hillas}
\end{figure*}
\begin{deluxetable*}{lcccccc}
\tablewidth{0pt}
\tablecaption{Typical values for acceleration region parameters and energy released by candidate sources.}
\tablehead{
\colhead{Candidate} & \colhead{R} & \colhead{B} & \colhead{Observed Luminosity} & \colhead{Duration} & \colhead{Event rate density} & Energy release\\
 & \colhead{[cm]} & \colhead{[Gauss]} & \colhead{[erg s$^{-1}$]} & \colhead{[s]} & \colhead{[Gpc$^{-3}$ yr$^{-1}$]} & \colhead{[erg]}
}
\startdata
HL-LGRB & $10^{13}$ - $10^{15}$ & $10^5$ - $10^7$ & $10^{50}$ & $\sim$20 & $\sim1$ & $2\times 10^{51}$\\
LL-LGRB & $10^{13}$ - $10^{15}$ & $10^2$ - $10^4$ & $5\times10^{46}$ & $\sim$20 & $\sim10^2$ & $10^{48}$\\
sGRB & \multicolumn{2}{c}{(same as HL-LGRB)} & $10^{50}$ & 0.3 & $\sim 1$ & $3\times10^{49}$\\
Jetted TDE & $10^{15}$ - $10^{17}$ & $0.1$ - $10$ & $10^{47}$ & $10^{6}$ & 0.03 &$10^{53}$\\
Non-jetted TDE & $10^{15}$ - $10^{17}$ & $0.01$ - $2$ & $10^{43}$ & $10^{6}$ & $5\times10^{2}$ & $10^{49}$\\
SLSNe - Shock interaction & $10^{16}$ - $10^{17}$ & 0.1 - 1 & $10^{44}$ & $10^{6}$ & 40 & $10^{49}$ - $10^{51}$\\
SLSNe - magnetar-powered & $10^6$ - $10^8$ & $10^{13}$ - $10^{15}$ & $10^{44}$ & $10^{6}$ & $\sim10^3$ & $10^{49}$ - $10^{51}$
\enddata
\tablecomments{For different subclasses of GRBs, the isotropic $\gamma$-ray energy release is presented. The acceleration radius $R=\Gamma R^\prime$ for GRBs and the magnetic field $B$ are found from \cite{Zhang:2003uk, Piran:2005qu, Murase:2006mm, Murase:2008mr}. We consider the upper range of values presented in the literature. For TDEs, the isotropic X-ray luminosity averaged over the first few weeks is considered. For jetted TDEs, we use the inferred value range from \cite{Kumar:2013apa, Duran:2013xoa, Yuan:2016zhk}, although it has been proposed that a pre-existing accretion disk may contain much higher magnetic flux \citep[see, e.g.,][]{Kelley:2014tga}. For non-jetted TDEs, the flare is in soft X-rays ($L_X/L_{\rm opt}\sim 0.1$), and no $\gamma$-ray emission can be observed. The values for non-jetted TDEs are based on radio observations of AT2019dsg \citep{Stein:2020xhk, Cendes:2021bvp}. Auchettl et al.~\citep{Auchettl:2016qfa} reported $L_{90} \sim 10^{44}$ and $10^{42}$ erg s$^{-1}$ for thermal and non-thermal TDEs, respectively, with $T_{90} \sim 10^{7}$ s, implying more stringent limits. For shock interaction-powered SLSNe, the majority of the emission is in the optical range \citep{Moriya:2018sig}, and the average event rate for Hydrogen-poor Type-I is $\sim 40$ \citep{Quimby:2013jb, Nicholl:2017slv}. We use the post-shock values for model parameters given in \cite{Margutti:2023ypj} inferred from the radio emission of Type-I SLSNe 2017ens. The event rate density of newly born magnetars associated with SLSNe is considered to be $\sim1\%$ of the core-collapse supernova rate \citep{Fang:2018hjp}. We assume $\lesssim10\%$ of the magnetar rotational energy goes into electromagnetic emission. The typical neutron star radius is $10^6$ cm. However, unipolar induction may occur close to the equatorial plane of the star in the wind outside the light cylinder \citep{Arons:2002yj, Fang_2012}.
}
\label{tab:sources}
\end{deluxetable*}

In the absence of a confirmed source association, we explore various transient candidates and test their viability against the neutrino energy injection rate
\begin{equation} 
\dot\varepsilon_\nu (E_\nu) = 4\pi {E_\nu^2 \Phi_\nu (E_\nu) 4\pi d_L^2}/{V_c(z)} \ , 
\label{eqn:nu_pt}
\end{equation}
where $E_\nu^2 \Phi_\nu (E_\nu)$ is the neutrino flux reported by KM3NeT, 
$d_L$ is the luminosity distance to the source and 
$V_c(z)$ is the comoving volume at redshift $z$. 


Efficient cosmic-ray acceleration requires the accelerator size to be larger than the Larmor radius in the comoving frame $R^\prime$, given by the Hillas condition \citep{Hillas_1984}, $E_{\rm max}\lesssim \eta^{-1}Ze\beta B R^\prime \Gamma$, which can be rewritten as
\begin{equation}
    BR^\prime \gtrsim 3.33 \times 10^{17} \eta \beta^{-1} \Gamma^{-1} E_{20} \ ,
\end{equation}
where $E_{20}=E_{\rm max}/10^{20}$ eV. For maximum acceleration efficiency, $\eta \simeq 1$ in the Bohm limit. Phenomenological models suggest a bulk Lorentz factor $\Gamma \sim 100$--$1000$ for luminous GRBs, which launch highly relativistic jets capable of accelerating cosmic rays in internal or external shocks. Jetted TDE flares, observed in optical, UV, and X-rays, typically exhibit $\Gamma \sim 10$.
However, jetted TDEs are extremely rare, interpreted as being powered by rapid accretion onto a supermassive black hole, with luminous, short-lived X-ray jets. Non-jetted TDEs and SLSNe lack relativistic outflows, possibly due to jet choking in stellar envelopes, and their luminosities peak in the optical/UV wavelengths. The latter can be powered by the interaction of the shock ejecta with dense circumstellar material. In magnetar-powered SLSNe, particles may be accelerated in relativistic winds acting as linear accelerators with $\Gamma \sim r/R^\prime$ \citep{Arons:2002yj}. These winds can disrupt the envelope, enabling cosmic-ray escape. Although magnetars may also power GRBs \citep{Zhang:2000wx}, we focus on millisecond magnetars as a distinct source class formed through core-collapse \citep{Murase:2009pg}, which are associated with SLSNe and fast blue optical transients.

We show the Hillas diagram on the left panel of Fig.~\ref{fig:hillas}. It is seen that shock-interacting SLSNe and non-jetted TDEs are disfavored as accelerators of UHECRs up to $\approx4$ EeV. The right panel of Fig.~\ref{fig:hillas} shows the isotropic energy release against the event rate density, compared to an order of magnitude estimate of the expected electromagnetic emission associated with the KM3NeT event. We assume that the optical depth for $p\gamma$ interaction is $\tau_{p\gamma}\gg1$ and hence the efficiency $f_{p\gamma}\sim 1$, which gives the maximum flux. Thus, for $\epsilon_e\approx0.1$, the energy in electromagnetic radiation is approximately an order of magnitude lower than in neutrinos. This conservative approach further disfavors choked-jet LL-GRBs and sGRBs as neutrino sources. The values used and the references are summarized in Table~\ref{tab:sources}.

The energy injection rates for the joint fit and KM3NeT-only fit flux levels are shown by the blue and orange curves in Fig.~\ref{fig:point_source}, respectively. The shaded bands indicate the flux uncertainty from KM3NeT data. This single-event estimate sets a lower bound on the energy injection rate from a transient source. Thus, consistency with the observed neutrino flux requires an adequate event rate, baryon loading, and efficient cosmic-ray interactions. We discuss some of the plausible candidate sources and their neutrino energy budget below.

\subsection{Gamma-ray bursts}

Gamma-ray bursts (GRB), with their relativistic jets capable of accelerating cosmic rays to ultrahigh energies, are a suitable candidate class for the source of this neutrino event. We compare the required neutrino energy injection rate with that from populations of short GRBs (sGRBs), high-luminosity long GRBs (HL-LGRBs), and low-luminosity long GRBs (LL-LGRBs). We consider the neutrino energy budget to be $\sim 10$ times that of the isotropic energy release in $\gamma$ rays as an optimistic upper limit (requiring $f_{p\gamma}=1$) since the fraction of shock energy in non-thermal electrons is $\epsilon_e\lesssim 0.1$ in the case of GRBs. The observed local event rate density of HL-LGRBs above an isotropic luminosity of $10^{50}$ erg s$^{-1}$  is $\sim1$ Gpc$^{-3}$ yr$^{-1}$, while that for LL-LGRBs above $5\times10^{46}$ erg s$^{-1}$ is $\sim10^2$ Gpc$^{-3}$ yr$^{-1}$ \citep{Sun:2015bda}. A typical duration of $\sim 20$~s yields the $\gamma$-ray energy released in the GRB blastwave.  Given the high energy of the neutrino event, a relativistic jet is a plausible origin, though not all compact object mergers produce jet breakouts. The local event rate density of sGRBs is uncertain and estimated to be about $\sim 0.5-3$ Gpc$^{-3}$ yr$^{-1}$ at above $10^{50}$ erg s$^{-1}$ \citep{Wanderman:2014eza, Sun:2015bda}. In this case, we consider the rate density to be $\mathcal{O}\sim1$ Gpc$^{-3}$ yr$^{-1}$ and a typical duration of 0.3~s to calculate the energy injection rate.

\begin{figure}
    \centering
    \includegraphics[width=0.48\textwidth]{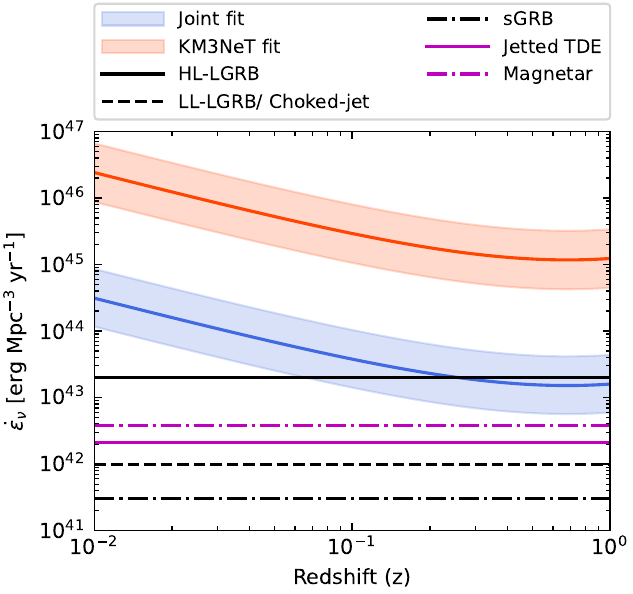}
    \caption{Local differential neutrino energy injection rate for the joint fit (blue) and KM3NeT-fit flux levels (red). The shaded regions correspond to the uncertainty in flux measurement. The energy injection rates for short GRBs, low-luminosity long GRBs, and high-luminosity long GRBs are shown for comparison, using characteristic values of injected energies and rate densities.}
    \label{fig:point_source}
\end{figure}

The energy injection rates for all three classes of GRBs are shown in Fig.~\ref{fig:point_source}. The observed event rate density depends on redshift and luminosity.
At low redshifts, the source evolution can be neglected. The neutrino energy budget inferred from the population of HL-LGRBs becomes comparable to that required to explain the joint-fit flux level at redshifts $z\gtrsim0.2$. This raises the possibility that the cumulative contribution from HL-LGRBs could account for the KM3NeT event. However, the lack of a coincident luminous GRB detection by $\gamma$-ray telescopes challenges this interpretation. If the source is a GRB, it is likely from the low-luminosity population (LL-GRB), which is more numerous but harder to detect via electromagnetic counterparts. But LL-GRBs and sGRBs are unable to account for the energy budget of the observed neutrino event unless a significantly larger energy is released in neutrinos compared to $\gamma$ rays in these transients.

\subsection{Tidal Disruption Events}

In TDEs, part of the debris falls into the black hole from an accretion disk, resulting in a flare in the UV or X-ray band \citep{Hills_1975, Frank_1976}. Depending on the accretion rate, a relativistic jet may be launched. 
It should be noted that among $\sim 100$ TDEs observed so far, only four are clearly known to have launched successful jets \citep{Stone:2020vdg}. 
Non-jetted TDEs, which have a much higher event rate, lack high-energy $\gamma$-ray emission and may produce a neutrino spectrum peaking at sub-PeV energies, depending on the production site \citep{Murase:2020lnu}. However, for non-jetted TDEs, accelerating UHECRs is challenging. 

Jetted TDEs are rare, but may accelerate UHECRs in jets.
In the case of GRBs, the bulk Lorentz factor is high, $\Gamma \sim 100$–$1000$. In contrast, TDE jets are expected to have $\Gamma \sim 10$, as they are not strongly confined by ambient gas \citep{Shcherbakov_2013}.
The duration of TDEs, $\Delta t\sim0.1$ - 1 yr, is much longer than GRBs. 
Jetted TDE like Swift J1644+57 triggered the Swift-BAT telescope for a total of $t_d=1.6\times 10^3$~s during 3 days of intense flaring with peak X-ray luminosity measured by Swift-XRT $L_{\rm peak} \sim 4.3\times 10^{48}$ erg s$^{-1}$ \citep{Burrows_2011}. The corresponding isotropic equivalent X-ray energy release is $L_{\rm peak} t_d\approx7.1\times 10^{51}$~ erg. Neutrino production up to tens of PeV energies has been proposed from such sources \citep{Wang_2011, Guepin:2017abw}. The event rate density of jetted Swift TDEs is $\approx 0.02-0.03$ Gpc$^{-3}$ yr$^{-1}$ for $L_{\rm peak} \sim 10^{48}$ erg s$^{-1}$ \citep{Sun:2015bda, Andreoni:2022afu}. The neutrino energy budget, assuming it is $\sim$10 times the isotropic X-ray energy release, is shown in Fig.~\ref{fig:point_source} to be much lower than the joint-fit flux level.

\subsection{Choked jets or failed GRBs}

Shock acceleration of cosmic rays is possible in jets that are stalled within the dense outer envelope of the progenitor star, near the photosphere \citep{Meszaros:2001ms, Zhu:2021mqc}. The jet deposits its energy entirely into the stellar envelope, and the breakout of the resulting mildly relativistic shock gives rise to an LL-GRB \citep{Nakar:2015tma}. Choked jets can also account for the connection between transrelativistic core-collapse supernovae of type Ib/c with failed GRBs. Cosmic rays up to $\sim$2–3~PeV in the comoving frame can be accelerated in internal shocks \citep{Razzaque:2004yv, Razzaque:2005bh} or collimation shocks \citep{Murase:2013ffa}. Although GeV–TeV $\gamma$-ray dark, choked jets can produce neutrinos \citep{Senno:2015tsn}. Owing to the quasi-isotropic shock breakout, neutrinos from this phase are detectable by off-axis observers without an electromagnetic counterpart.

Due to $\Gamma \lesssim 10$, cosmic-ray acceleration to EeV energies is challenging, as radiation-mediated shocks form prior to jet breakout. To allow for collisionless shocks and particle acceleration, the kinetic luminosity in the jet is required to be $L<10^{47}$ erg s$^{-1}$ \citep{Murase:2013ffa}. The dominant energy loss mechanism is $p\gamma$ interactions with photons produced both in internal shocks and at the jet head. The low $\Gamma$ and the cosmic-ray loading parameter constrain the resulting TeV neutrino flux. The interaction efficiency can reach $f_{p\gamma} \sim 1$, making choked jets a viable candidate for the orphan neutrino event KM3-230213A, given the uncertainty in the reconstructed neutrino energy. Assuming a jet breakout time of $\sim17$~s \citep{Senno:2015tsn} and a local LL-GRB rate of $\sim 10^2$ Gpc$^{-3}$ yr$^{-1}$ at $L_\gamma = 5 \times 10^{46}$ erg s$^{-1}$, we recover the LL-LGRB neutrino energy budget derived earlier, adopting a neutrino output $\sim10$ times the isotropic-equivalent $\gamma$-ray luminosity.

\subsection{Magnetar-powered SLSNe}

Magnetars are ultramagnetized neutron stars with surface magnetic fields of $10^{14}-10^{15}$\,G \citep{Duncan_1992,Turolla:2015mwa} typically formed in the core-collapse of massive stars with progenitor masses around $10-40$~$M_\odot$ \citep{Ferrario:2006ib}. They can also be produced as remnants of binary neutron star mergers if the maximum non-rotating neutron star mass exceeds $\sim2.2-2.4$~$M_\odot$ for a stiff equation of state, allowing the remnant to avoid prompt collapse \citep{Margalit:2017dij, Shibata:2017xdx}. Magnetars have thus been proposed as central engines of GRBs, superluminous supernovae, and other energetic transients \citep{Dai:2006hj, Metzger_2011}.

Newly born magnetars rotating rapidly with a period of about $\sim 1$~ms can store up to $\sim 10^{52}$–$10^{53}$ erg in rotational energy, which can be extracted via a strong dipole magnetic field \citep{Thompson:2004wi}. Cosmic-ray acceleration may occur in magnetic reconnection layers, or at the termination shock of the magnetar wind nebula. In the merger scenario, they can interact with the X-ray, optical and UV photons producing neutrinos \citep{Metzger:2013cha, Fang:2017tla}. For a magnetar-powered superluminous supernovae rate of $\sim1\%$ of the core-collapse supernovae, $600-1200$ Gpc$^{-3}$ yr$^{-1}$ and assuming a fraction $\eta=0.1$ of the rotational energy goes into cosmic rays, the cosmic-ray energy budget is estimated to be $(0.3-6)\times 10^{44}$ erg Mpc$^{-3}$ yr$^{-1}$ for a spin period of $2-10$~ms \citep{Fang:2018hjp}. The corresponding neutrino energy injection rate, $\dot\varepsilon_\nu \approx (3/8)f_{p\gamma}\dot\varepsilon_{\rm CR}$, is compared with that required for KM3-230213A, assuming an efficiency $f_{p\gamma}\sim 0.1$ and CR energy budget $\sim 1\times10^{44}$ erg Mpc$^{-3}$ yr$^{-1}$.

\subsection{Off-axis GRB jets}

For structured jets or arbitrary viewing angles, the observed flux can be significantly reduced due to diminished Doppler boosting, as $\gamma$ rays are strongly beamed when the viewing angle $\theta_{\rm ob} < \theta_{\rm jet}$. However, off-axis GRBs with $\theta_{\rm ob} > \theta_{\rm jet}$ can produce high-energy neutrinos without accompanying electromagnetic counterparts, provided that cosmic rays are scattered by the circumburst magnetic field or a large-scale magnetized structure into the observer’s line of sight, enabling $p\gamma$ interactions outside the jet cone.

The observed flux from an off-axis homogeneous fireball can be obtained by rescaling $F_{\rm off} (\theta_{\rm ob})= F_{\rm on}(0)\theta^2/8\Gamma D^3$ at times earlier than the break in afterglow lightcurve, where $D=[\Gamma(1-\bm{\beta\cdot \hat{n}_{\rm ob}})]$ is the Doppler factor and $\theta$ is the effective angular size contributing to the observed flux \citep{Rossi:2001pk, Ahlers:2019fwz}. For relativistic jets near the break time, $\theta\simeq\Gamma^{-1}$. The relative event rate of off-axis jets can be obtained using the factor $(1-\cos\theta_{\rm ob})/(1-\cos\theta_{\rm jet})$ to account for the geometric suppression of detectable events due to jet collimation. For $\theta_{\rm ob}\approx 10^\circ$, $\theta_{\rm jet}\approx5^\circ$, and $\Gamma\sim100$, the observed $\gamma$-ray flux is around $\sim10^{-7}$ times lower than HL-LGRBs. Assuming the neutrino energy scales as $\sim10$ times the $\gamma$-ray energy release, the KM3NeT observation is indicative of a rare, stochastic event, if arising from an off-axis jet.

\subsection{General remarks}

For a characteristic source with an isotropic neutrino energy budget of $10^{50}$ ergs, the joint-fit flux level would imply a local event rate density of $\sim400$ Gpc$^{-3}$ yr$^{-1}$. 
While the Hillas criterion allows HL-LGRBs, magnetars, and jetted TDEs as plausible candidates, the flux level observed by KM3NeT remains difficult to explain with these sources, as shown in Fig.~\ref{fig:point_source}. HL-LGRBs can account for the joint-fit flux level if $f_{p\gamma} \approx 1$, but lower values would put them in tension with the KM3NeT observation.
%
%
The production of neutrinos is accompanied by a comparable flux of $\gamma$-rays. It should be noted that currently operating $\gamma$-ray detectors have limited sky coverage, and the proximity of the neutrino event to the Galactic plane further hinders source identification due to foreground Galactic emission obscuring potential counterparts.
Neutrino luminosity and energy fall sharply in off-axis directions for a homogeneous jet. A viewing angle slightly larger than the jet opening angle ($\theta_{\rm ob} \gtrsim \theta_{\rm jet}$) may produce a single neutrino event without detectable $\gamma$-rays, but the flux is insufficient to account for one event during KM3NeT’s operation time.
Alternatively, the event may be associated with an unrecognized class of transients that can accelerate UHECRs up to a few times $10^{18}$ eV, exhibit a suppressed $\gamma$-ray flux due to cascade emission in a dense environment near the source or are absorbed by pair production, and occur with a rate density compatible with the inferred energy budget.



\section{Diffuse astrophysical emission\label{sec:diffuse}}

\begin{figure*}
    \centering
    \includegraphics[width=0.97\textwidth]{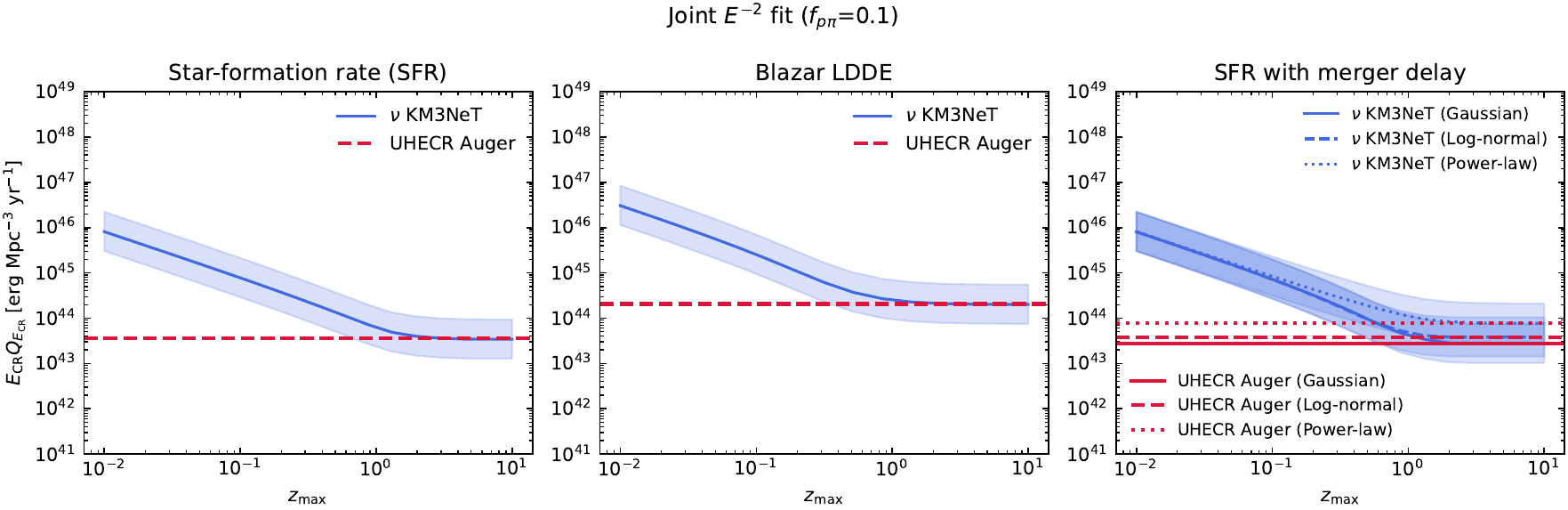}
    \caption{The blue curve shows the local differential cosmic-ray energy injection rate required for the UHE$\nu$ detection by KM3NeT as a function of the maximum redshift of the source population. To compute this rate, we convert the single-flavor flux $E_\nu^2 \Phi^{\rm 1f}_{\nu + {\bar\nu}}(E) = (7.5^{+13.1}_{-4.7}) \times 10^{-10}~\mathrm{GeV\,cm^{-2}\,s^{-1}\,sr^{-1}}$ reported by KM3NeT in the 72 PeV--2.6 EeV range to all-flavor flux. The value corresponds to the joint KM3NeT, IceCube, and Auger analysis. We assume the efficiency of converting UHECRs to neutrinos at this energy is $f_{\rm p,\pi}=0.1$ in the sources.
    The blue-shaded region corresponds to the neutrino flux uncertainty. The red dashed curve corresponds to the local differential cosmic-ray energy injection rate obtained using Auger UHECR flux at $\gtrsim 4$~EeV for a fixed maximum redshift $z_{\rm max} = 3$.
    We explore the redshift evolution of sources given by star-formation rate (\textit{left panel}), a luminosity-dependent density evolution function for blazars (\textit{middle panel}), and star-formation rate with time delay due to compact binary merger events (\textit{right panel}). For the latter, we use the redshift evolution function given by Gaussian delay, log-normal delay, and power-law delay. The required UHECR energy injection rate derived from KM3NeT observation exceeds the allowed limit from Auger measurements if the neutrino population is bound to $z\lesssim1$}.
    \label{fig:diff1}
\end{figure*}
\begin{figure*}
    \centering
    \includegraphics[width=0.97\textwidth]{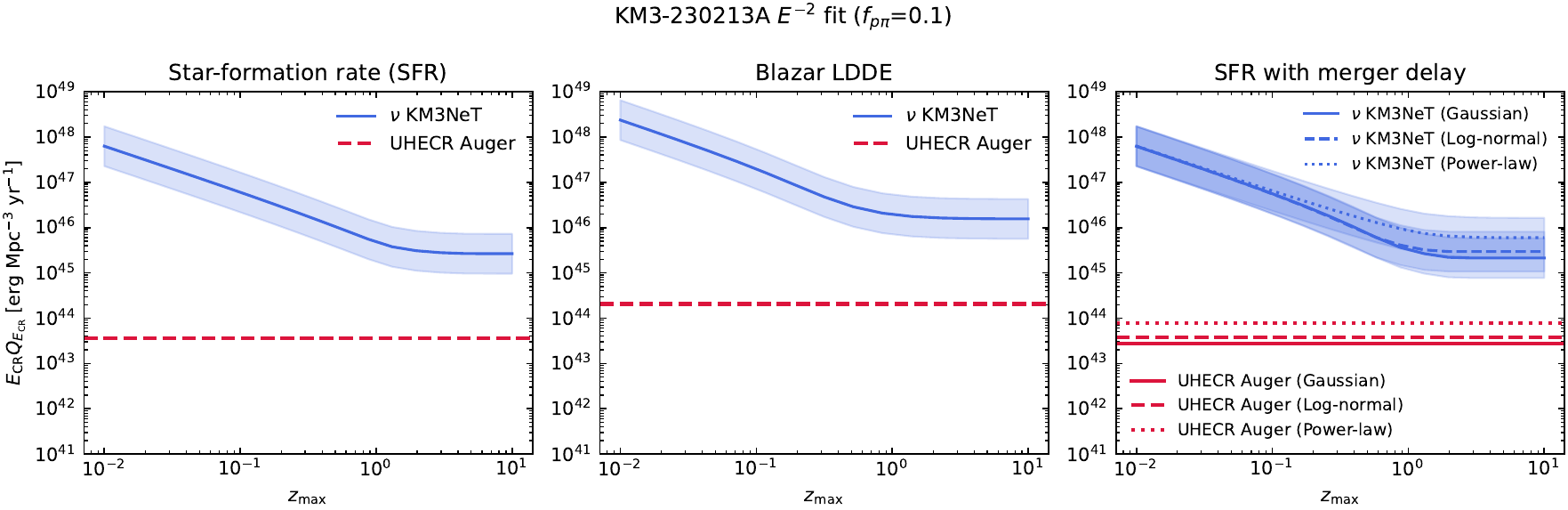}
    \caption{Same as Fig.~\ref{fig:diff1} but for a 
    single-flavor flux of $E^2 \Phi^{\rm 1f}_{\nu + {\bar\nu}}(E) = (5.8^{+10.1}_{-3.7}) \times 10^{-8}~\mathrm{GeV\,cm^{-2}\,s^{-1}\,sr^{-1}}$
    corresponding to the KM3NeT-only fit to the neutrino event. In this case, the required UHECR energy injection rate for the neutrino flux exceeds the Auger limit at all redshifts.}
    \label{fig:diff2}
\end{figure*}

If the inferred neutrino flux is attributable to an unresolved population of sources contributing to a diffuse background, then the flux can be written as \citep[see, e.g.,][]{Stecker_1993, Murase:2007ar, Murase:2012df, Murase:2010gj}
\begin{equation}
    E_\nu^2\Phi_\nu^{\rm diff}(E_\nu) = \int dz \frac{dV}{dz} \frac{1}{4\pi d_L^2}E_\nu^2\frac{d\dot n}{dE}(z) \ ,
    \label{eqn:source}
\end{equation}
%
%
where $dV/dz$ is the cosmological volume element and $E_\nu^2{d\dot n}/{dE_\nu}(z)$ is the observed neutrino spectrum due to injections at redshift $z$. The total neutrino energy injection rate is given by $Q_\nu(z)= \int dE_\nu^\prime E_\nu^\prime{d\dot n^\prime}/{dE_{\nu^\prime}}(z)$. The local energy injection rate is $Q_\nu\equiv Q_\nu(z=0)$ and $dV/dz = cd_L^2d\Omega/H_0(1+z)^2E(z)$, where $d_L$ is the luminosity distance at $z$ and $H_0$ is the Hubble constant. $E(z)=H(z)/H_0$ is the dimensionless Hubble parameter at redshift $z$. Rewriting Eq.~(\ref{eqn:source}) with these substitutions, we have
\begin{equation}
    E_\nu^2 \Phi^{\rm diff}_\nu(E_\nu) = \frac{c}{4\pi}(E_\nu Q_{E_\nu})\int \frac{dz}{1+z}\bigg| \frac{dt}{dz}\bigg| f(z) \ ,
    \label{eqn:budget}
\end{equation}
where $E_\nu Q_{E_\nu}$ is the local differential energy input of neutrinos and $f(z)$ is the redshift evolution function. The cosmological line element is given by $
| {dt}/{dz}|^{-1} = {H_0(1+z) \sqrt{(1+z)^3\Omega_m + \Omega_\Lambda}}$. We consider $H_0=67.3$ km s$^{-1}$ Mpc$^{-1}$, $\Omega_m=0.315$ and $\Omega_\Lambda=0.685$ \citep{Planck:2018vyg}. For astrophysical sources producing neutrinos via $p\gamma$ interactions, the corresponding cosmic-ray energy injection rate is calculated as \citep{Waxman:1997ti} 
\begin{equation}
    E_\nu Q_{E_\nu} = \frac{3}{8}\min[1, f_{p,\pi}]E_{\rm CR}Q^{\rm KM3}_{E_{\rm CR}} \ ,
\end{equation}
where the photomeson production efficiency is given by $f_{p,\pi}$, that depends on the source class. 
For a power-law target photon spectrum, $ {dn}/{d\varepsilon} = n_0 \left({\varepsilon}/{\varepsilon_0}\right)^{-\beta} (\beta \gtrsim 1), $ we have
\begin{equation}
    f_{p,\pi} \approx \frac{2\varepsilon_0 n_0}{1+\beta} c t_{\rm int} \sigma_{\Delta} 
    \frac{\Delta \bar{\varepsilon}_{\Delta}}{\bar{\varepsilon}_{\Delta}} 
    \left( \frac{E_{\rm cr}}{E_{\rm cr}^0} \right)^{\beta - 1},
\end{equation}
where $ t_{\rm int} $ is the interaction time between CRs and photons, 
$\sigma_{\Delta} \sim 5 \times 10^{-28} \, \text{cm}^2 $, 
$ \bar{\varepsilon}_{\Delta} \sim 0.3 \, \text{GeV} $, 
$ \Delta \bar{\varepsilon}_{\Delta} \sim 0.2 \, \text{GeV} $, 
$E_{\rm cr}^0 \approx 0.5 \delta^2 \bar{\varepsilon}_{\Delta} m_p c^2 / \varepsilon_0,
$
and $ \delta $ is the Doppler factor of the CR source. 
We expect $f_{p,\pi} \sim 0.01 - 0.1$ for GRB prompt emission. In the limit of semi-transparent sources, $f_{p, \pi} \rightarrow1$ yields the Waxman-Bahcall bound \citep{Waxman:1998yy}. However, we assume a conservative limit of $f_{p,\pi}=0.1$ for our analysis.

The UHECR energy injection rate derived from the neutrino flux depends on the distribution of sources in redshift and the assumed maximum redshift. The Spitzer space telescope constrains the cosmic star formation history up to $z\sim 1$ \citep{Coward:2007jw}. At higher redshifts, a plateau in the range $z\sim1-4$ and a steep decrease at $z\gtrsim4$ is expected \citep[see, e.g.,][]{Hopkins:2006bw, Yuksel:2008cu},
\begin{align}
    \psi_{\rm SFR}(z) \propto
    \begin{cases} 
        (1+z)^{3.44}, & z < 0.97 \\
        10^{1.09} (1+z)^{-0.26}, & 0.97 < z < 4.48 \\ 
        10^{6.66} (1+z)^{-7.8}, & z > 4.48
    \end{cases}
\end{align}
In the case of blazars, the LDDE function is given in \cite{Ajello_2012, Ajello:2013lka} and $f(z)$ in Eq.~(\ref{eqn:budget}) is replaced by $\int dL_{100} f(L_{100},z)$, where $L_{100}$ is the $\gamma$-ray luminosity between 100 MeV to 100 GeV. Integrating over suitable parameter ranges gives the distribution of resolved and unresolved blazars in redshift-luminosity space \citep[see, e.g.,][]{Palladino:2018lov, Das:2020hev}. We add the contribution from BL Lacertae objects (BL Lacs) and flat spectrum radio quasars (FSRQs) to estimate the total energy injection rate.

The merger event rate density of binary neutron stars (BNS) and black hole–neutron star (BH–NS) systems, $\dot\rho(z)$, at redshift $z$, can be connected to the cosmological star formation rate density by accounting for the merger delay time from the formation of the binary system and the orbital decay timescale due to gravitational wave radiation. We use the functional form presented in the Appendix of \cite{Zhu:2020ffa} for $f(z)$, assuming Gaussian \citep{Virgili_2011}, log-normal, and power-law delay time distributions \citep{Wanderman:2014eza}. These distributions are based on NS-NS mergers, originally used to model the observed redshift distribution of short GRBs. NS–BH mergers are also believed to follow a broadly similar distribution.

%
%
%
%

The energy loss length of protons at 4 EeV is a few Gpc \citep{Dermer:2008cy}, with Bethe–Heitler pair production on the extragalactic background light (infrared, optical, and ultraviolet photons) being the primary loss process, though its effect at this energy is negligible.
Hence, using the same equation as Eq.~(\ref{eqn:budget}), we calculate the local differential energy injection rate $E_{\rm CR}Q^{\rm Auger}_{E_{\rm CR}}$ from Auger spectrum data using a fixed value of maximum redshift $z_{\rm max}=3$ (red dashed curve in Figs.~\ref{fig:diff1} and \ref{fig:diff2}). The left-hand side of the equation is replaced by the latest UHECR flux measurement by Auger \citep{PierreAuger:2023kgv}.
The contribution of UHECRs beyond this value is negligible since the source evolution falls off sharply.  
In Figs.~\ref{fig:diff1} and \ref{fig:diff2}, we present the variation of the UHECR energy budget with maximum redshift, deduced from KM3NeT neutrino detection (blue curve) for the joint fit and KM3NeT-only fit, respectively. The blue-shaded region corresponds to the flux uncertainty provided by KM3NeT \citep{KM3NeT:2025ccp}. The left, middle, and right panels show the redshift evolution for SFR, blazar LDDE, and SFR with merger delay, respectively. 
\begin{figure*}
    \centering
    \includegraphics[width=0.34\linewidth]{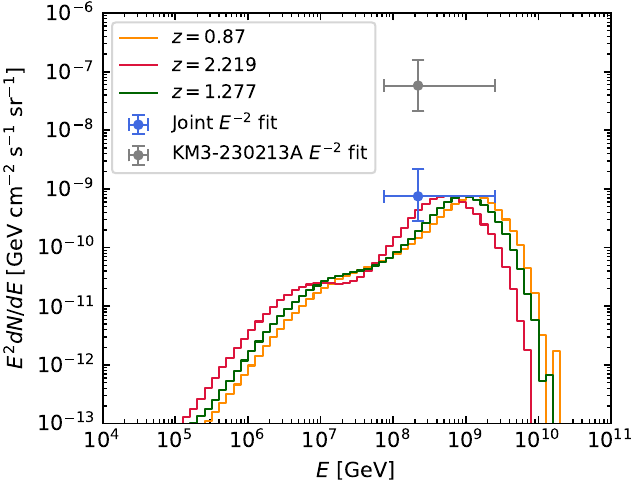}%
    \includegraphics[width=0.33\textwidth]{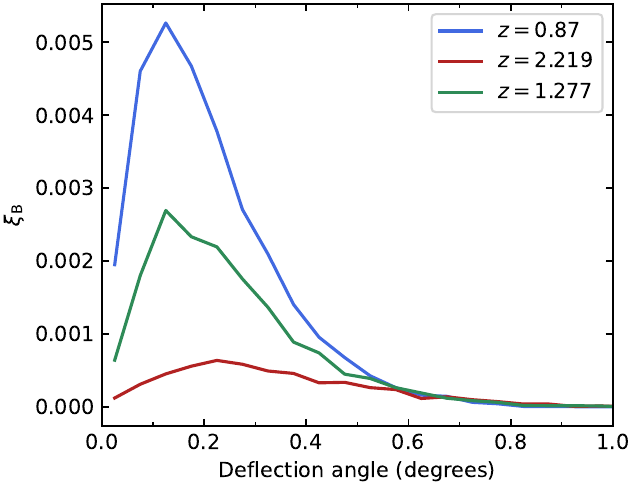}%
    \includegraphics[width=0.33\textwidth]{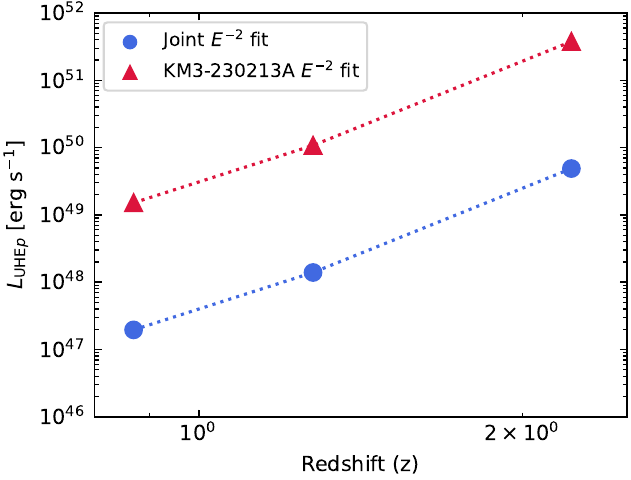}
    \caption{\textit{Left:} The single-flavor cosmogenic neutrino flux due to cosmological propagation of UHE protons injected from candidate blazar sources at various redshifts. The peak flux is normalized to the obtained flux at 220 PeV from the joint $E^{-2}$ fit. \textit{Middle:} The survival fraction of UHECRs as a function of angular deflection bins of 0.05$^\circ$ with respect to the line of sight. \textit{Right}: The corresponding luminosity requirement in UHE$p$ from these sources to explain the observed neutrino event converted to all-flavor flux for each of the fits by KM3NeT collaboration.}
    \label{fig:cosmogenic}
\end{figure*}

In Fig.~\ref{fig:diff1}, the UHECR energy injection rate required to explain the observed neutrino in the joint fit becomes consistent with the Auger flux limits when the maximum redshift of the source population satisfies $z_{\rm max}\gtrsim1$ for all source evolution cases. 
For source contribution from $z_{\rm max}<1$, the neutrino flux violates the Auger limit on the energy injection rate. 
However, Fig.~\ref{fig:diff2} shows that the KM3NeT-only fit is difficult to reconcile with the constraints from Auger. This suggests that an interpretation based on diffuse astrophysical emission is tenable, given the flux level obtained from the joint fit.
The exact value of $p\gamma$ interaction efficiency depends on the target photon energy, flux, and also the escape of cosmic rays from the source environment. The opacity of $p\gamma$ interactions can be much lower than that considered in this section. For $f_{p\gamma} < 0.1$, even the joint fit flux level fails to comply with the allowed cosmic-ray energy budget. On the other hand, the KM3NeT-only fit remains in tension with the cosmic-ray energy budget even in the semi-transparent limit of $f_{p\gamma}\sim 1$.

\section{Cosmogenic neutrino from blazars\label{sec:cosmogenic}}

Among 17 candidate blazars within $3^\circ$ (99\% C.L.) of the reconstructed KM3NeT-ARCA event reported in \citet{KM3NeT:2025bxl}, only three sources \#6, \#7, and \#8 have known redshifts. These are
5BZCAT sources classified as FSRQs  \citep{Massaro:2015nia}.
Their
redshifts are $z=0.87$, 2.219, and 1.277, respectively.
We study the possibility of the neutrino event originating from the line-of-sight interaction of UHECRs injected by blazars \citep[see, e.g.,][]{Essey:2010er, Das2022A&A...658L...6D}. 
We simulate 1-D UHECR propagation in the extragalactic medium using the Monte Carlo framework \textsc{CRPropa}~3.2 \citep{AlvesBatista:2016vpy, AlvesBatista:2022vem}, accounting for all relevant energy loss processes, photopion production, and Bethe–Heitler pair production with the cosmic microwave background and extragalactic background light, adiabatic losses due to cosmic expansion, and $\beta$-decay of secondary neutrons.
The left panel in Fig.~\ref{fig:cosmogenic} shows the all-flavor cosmogenic neutrino fluxes, assuming an injection spectrum $dN/dE\propto E^{-2}$ between 0.1 and 100 EeV. The peak of the simulated spectrum is normalized to the measured flux by KM3NeT, corresponding to the joint fit. 

We perform 3-D simulations of UHECR deflection in a random turbulent extragalactic magnetic field defined over 0.1 to 5 Mpc length scale with an RMS field strength of $B=10^{-5}$~nG and a turbulence coherence length $\lambda_c\approx 1$~Mpc. Propagated events are recorded on the surface of an observer sphere of radius 1 Mpc centered on Earth. The middle panel in Fig.~\ref{fig:cosmogenic} shows the survival fraction $\xi_B$ of UHECRs as a function of deflection angle with respect to emission direction. The kinetic power in UHE protons required at the source is calculated using the expression \citep{Das:2019gtu},
\begin{equation}
L_{\rm UHEp} = \dfrac{2\pi d_L^2(1-\cos\theta_{\rm jet})}{\xi_B f_{\rm CR}} \int_{\epsilon_{\nu,\rm min}}^{\epsilon_{\nu,\rm max}} \epsilon_\nu\dfrac{dN}{d\epsilon_\nu dAdt}d\epsilon_\nu \label{eqn:luminosity} \ ,
\end{equation}
where $2\pi d_L^2(1-\cos\theta_{\rm jet})$ is the area of the spherical cap region subtended by the blazar jet at the distance of the observer. We consider typical values of jet opening angle $\theta_{\rm jet}\sim 0.1$ radians \citep{Pushkarev_09, Finke_19}. $f_{\rm CR}$ is the fraction of injected power that goes into secondary neutrinos, and the integration is over the cosmogenic flux. We consider the fraction of UHECRs within 0.3$^\circ$ of the initial emission direction, which is 10\% of the angular uncertainty. 

The right panel of Fig.~\ref{fig:cosmogenic} shows the required luminosity as a function of redshift, where red and blue points correspond to the flux levels for the KM3NeT-only fit and the joint fit, respectively. The values obtained are higher than the typical Eddington luminosity of the blazars, calculated using the expression $L_{\rm Edd}=10^{47}(M_{\rm BH}/10^9 M_\odot)$ erg s$^{-1}$ \citep{Falomo_03}, where $M_{\rm BH}$ is the mass of the central supermassive black hole. For $z\gtrsim 1$, the luminosity for both cases tends to be much higher than typical of AGNs. However, for an origin at $z\lesssim 1$, line-of-sight UHECR interactions cannot be ruled out for the joint fit flux level.

Modeling of the Auger spectrum indicates a Peter’s cycle composition, characterized by a rigidity-dependent cutoff ($\log_{10}(R_{\rm cut}/V)\sim 18.1 - 18.7$) and a hard injection spectral index. A combined fit to the energy spectrum and shower-depth distribution ($X_{\rm max}$ and $\sigma(X_{\rm max})$) measured by Auger reveals a proton-dominated composition near $4~\mathrm{EeV}$, based on post-LHC hadronic interaction models \citep[see, e.g.,][]{PierreAuger:2016use, PierreAuger:2024flk, AlvesBatista:2018zui, Das:2018ymz}. Heavier nuclei dominating above $\sim 30~\mathrm{EeV}$ must originate from the local Universe ($z \lesssim 0.5$) to avoid significant photodisintegration. Heavy nuclei from redshifts of these candidate blazars may disintegrate during propagation or inside the sources depending on $f_{p\gamma}$, producing secondary nucleons that contribute to the sub-ankle flux at $\lesssim 5\times 10^{18}$ eV and generate cosmogenic neutrinos peaking at $\sim 10^{17}$~eV. However, even at $\sim 100~\mathrm{EeV}$, $10-15$\% proton composition is allowed, and the resulting neutrino spectrum dominates over that from heavier nuclei at the KM3NeT neutrino energy, assuming UHECR sources up to $z_{\rm max}=1$ \citep[see, e.g.,][]{Das:2018ymz, Das:2020nvx}.

The IGMF uncertainty can largely impact the estimates of kinetic luminosity in blazar sources. For a Larmor radius $r_L$, the expected deflection $\theta_{\rm dfl}= {\sqrt{2d_c\lambda_c}}/{3r_L}$ for $z=0.87$ can be expressed as \citep{Dermer:2008cy, Murase:2011yw}
\begin{align}
    \theta_{\rm dfl} \simeq 0.13^\circ\, Z 
    &\left( \frac{d_c}{3~\mathrm{Gpc}} \right)^{1/2}
    \left( \frac{\lambda_c}{1~\mathrm{Mpc}} \right)^{1/2}  \nonumber \\
    &\times\left( \frac{B_{\rm EG}}{10^{-5}~\mathrm{nG}} \right)
    \left( \frac{E_{p}}{100~\mathrm{EeV}} \right)^{-1}\,.
    \label{eqn:defl}
\end{align}
Sources \#6, \#7, \#8 reported in \cite{KM3NeT:2025bxl} have an angular separation of $\approx 2.5^\circ$ from the neutrino event.
We adopt an optimistic value for the RMS strength, $B_{\rm EG} \sim 10^{-5}$~nG, given that current constraints allow values as low as $10^{-7}$~nG \citep{Neronov_2010}.
The deflection in Eq.~\ref{eqn:defl} can increase with higher magnetic field strength or lower UHE proton energy, potentially scattering UHECRs out of our line of sight for distant sources. Even if the EGMF is higher by an order of magnitude, the deflection remains smaller than both the jet opening angle of $\theta_{\rm jet} = 0.1$ radian and the value of $0.3^\circ$ used in calculating $f_{\rm CR}$ and thus, the source luminosity. However, the angular resolution of the KM3NeT detection ($\approx 3^\circ$ at 99\% C.L.) introduces additional uncertainty in interpreting magnetic deflection. Since the mean free path for $p\gamma$ interactions exceeds the size of the Galaxy, it is unlikely that KM3-230213A was produced by UHECR interactions within the Milky Way, and Galactic magnetic deflections can therefore be neglected. UHECR composition and uncertainties in the EGMF deflection limit our predictive capability; thus, the results presented here can be regarded as a benchmark reference scenario.

However, we note that the neutrino signal may also originate away from the line of sight to the source. UHECRs escaping their sources can undergo strong deflection in the surrounding large-scale structure, such as clusters of galaxies or cosmic filaments, where the EGMF strength can be $\gtrsim 1$ nG. The magnetized environment surrounding the source may vary significantly across different candidate classes. For example, TDEs are often found in post-starburst galaxies, while magnetars and GRBs may reside in star-forming regions associated with young stellar populations. These additional deflections can increase the required luminosity by introducing an angular offset between the source and the neutrino arrival direction. Moreover, the deflection of secondary leptons in the magnetized structure translates into an additional time delay of the cosmogenic $\gamma$-ray signal produced by the electromagnetic cascade initiated by UHECRs.

\section{Discussions \& Conclusions \label{sec:discussions}}

Based on the analysis presented in this article, the flux from the KM3NeT-only fit is difficult to reconcile with all three scenarios considered.
In the case of a transient point source, we show the energy injection rate of HL-LGRB, LL-LGRB, and sGRBs. Although the neutrino energy budget of the HL-LGRB population at $z\gtrsim 0.2$ is comparable to that inferred for the point source origin of the KM3NeT event, the non-detection of a luminous $\gamma$-ray source challenges this scenario. The near-horizontal trajectory of the KM3NeT event makes it difficult to identify electromagnetic counterparts. 
In addition, for $f_{p\gamma}<1$, HL-LGRBs fail to comply with the joint-fit flux level. However, the possibility of an isolated neutrino event from an off-axis jet, although an unlikely scenario, cannot be ruled out.
The combined data from KM3NeT, IceCube, and the Pierre Auger Observatory support the possibility of a source flare lasting $\lesssim 2$ years \citep{Neronov:2025jfj}. However, such neutrino-flaring sources are expected to be rare among known transient populations. 
Moreover, jetted TDEs and magnetars may accelerate UHECRs up to 4 EeV (see Fig.~\ref{fig:hillas}), but they cannot account for the neutrino energy injection rate required for this event (see Fig.~\ref{fig:point_source}). This could also indicate the existence of a previously unidentified class of transient source with a high local event rate density and which is $\gamma$-ray dark, i.e., high-energy $\gamma$-rays are absorbed within the source.

If the flux arises from a diffuse astrophysical emission, the source distribution is required to extend up to $z_{\rm max}\gtrsim 1$ to conform with the energy injection rate derived from Auger flux measurements. All the redshift evolution models considered here, viz., SFR, blazar LDDE, and SFR with merger delay, can account for the required UHECR energy injection rate for the joint fit.
We find that our choice of $p\gamma$ interaction efficiency, $f_{p\gamma} = 0.1$, serves as a limiting case for the joint fit. Lower values challenge the diffuse emission scenario based on Auger flux limit, while the KM3NeT-only fit remains disfavored even in the semi-transparent limit of $f_{p\gamma} \rightarrow 1$.
A diffuse neutrino flux, composed of astrophysical and cosmogenic components from UHECR sources distributed according to the star-formation rate (SFR), has been shown to explain the KM3NeT observations \citep{Muzio:2025gbr}. The cosmic-ray energy budget derived from the Auger flux at $\sim 4$~EeV is a conservative lower limit, assuming negligible interactions at this energy. With a mean free path of $\simeq 2$~Gpc ($z \approx 0.5$), energy losses are minimal for nearby sources but may become significant at higher redshifts due to evolving cosmic photon backgrounds.

Under the cosmogenic hypothesis, explaining the neutrino event through line-of-sight UHECR interactions from candidate sources at high redshifts requires a highly collimated UHECR beam. A flaring or transient event can inject UHECRs that propagate along our line of sight and interact near the source. Over cosmological distances, even weak extragalactic magnetic fields can significantly deflect UHECRs, substantially increasing the required proton luminosity. The KM3NeT collaboration has provided a list of potential blazar candidates lying within the angular uncertainty  \citep{KM3NeT:2025bxl}. For an RMS field strength of $10^{-5}$~nG, the necessary kinetic power in protons exceeds typical values expected from AGNs at $z\gtrsim 1$.
For the local cumulative number density of blazars $\sim10^{-6}$ Mpc$^{-3}$, the luminosity at $z\lesssim0.1$ is $\mathcal{O}\sim 10^{46}$ erg s$^{-1}$, 
which corresponds to powerful blazars with their synchrotron luminosity equivalent to the UHECR luminosity \citep{Razzaque:2011jc},
indicating the viability of AGN sources.  The luminosity could be much higher for a flaring source or a transient event. The typical isotropic $\gamma$-ray luminosity of GRBs can reach $10^{51}$-$10^{53}$ erg s$^{-1}$, which may explain the highest power required for KM3NeT-fit on the right panel of Fig.~\ref{fig:cosmogenic}, under the line-of-sight UHECR interaction scenario.  
 

The KM3-230213A event demonstrates the capability of next-generation neutrino telescopes to probe extreme energies and rare sources. Joint observations of neutrinos, $\gamma$-rays, and UHECRs are crucial for uncovering the origin of such events and constraining the physics of the most powerful cosmic accelerators.
While proposing new astrophysical sources is speculative, the observation by KM3NeT provides a compelling opportunity to test various alternative phenomenological models and candidate source classes.





\begin{acknowledgments}
    S.D. and S.X. acknowledge the support from NASA ATP award 80NSSC24K0896.
    B.Z. acknowledges NASA 80NSSC23M0104 for support; 
    S.R.\ was partially supported by a BRICS STI grant and by a NITheCS grant from the National Research Foundation, South Africa.
\end{acknowledgments}

\bibliography{sample631}{}
\bibliographystyle{aasjournal}



\end{document}